\documentclass[a4paper,11pt]{article}
\usepackage{pos}

\usepackage{lipsum}

\let\OLDthebibliography\thebibliography
\renewcommand\thebibliography[1]{
  \OLDthebibliography{#1}
  \setlength{\parskip}{1.4pt}
  \setlength{\itemsep}{0.6pt plus 0.5ex}
}

\usepackage{graphicx}
\usepackage{dcolumn}
\usepackage{bm}
\usepackage{blindtext}
\usepackage{xcolor}
\usepackage{amsthm}
\usepackage{amsmath}
\usepackage{hyperref}

\newcommand{\Bern}{Institute for Theoretical Physics, Albert Einstein Center for Fundamental Physics,\\University of Bern, Sidlerstrasse 5, CH-3012 Bern, Switzerland}
\newcommand{\hiskp}{HISKP (Theory), Rheinische Friedrich-Wilhelms-Universit\"at Bonn,\\Nussallee 14-16, 53115 Bonn, Germany}
\newcommand{\hpca}{High Performance Computing and Analytics Lab, Rheinische Friedrich-Wilhelms-Universit\"at Bonn,\\ Friedrich-Hirzebruch-Allee 8, 53115 Bonn, Germany}
\newcommand{\CyprusU}{Department of Physics, University of Cyprus, 20537 Nicosia, Cyprus}
\newcommand{\CyprusI}{Computation-based Science and Technology Research Center, The Cyprus Institute,\\20 Konstantinou Kavafi Street, 2121 Nicosia, Cyprus}

\newcommand{\Parma}{Dipartimento  di  Scienze  Matematiche,  Fisiche  e  Informatiche,  Universit\`a  di  Parma  and  INFN, Gruppo  Collegato  di  Parma,  Parco  Area  delle  Scienze  7/a  (Campus),  43124  Parma,  Italy}
\newcommand{\Romadue}{Dipartimento di Fisica and INFN, Universit\`a di Roma ``Tor Vergata",\\Via della Ricerca Scientifica 1, I-00133 Roma, Italy}
\newcommand{\Romatre}{Dipartimento di Matematica e Fisica, Universit\`a Roma Tre and INFN, Sezione di Roma Tre,\\Via della Vasca Navale 84, I-00146 Rome, Italy}
\newcommand{\RomatreINFN}{Istituto Nazionale di Fisica Nucleare, Sezione di Roma Tre,\\Via della Vasca Navale 84, I-00146 Rome, Italy}
\newcommand{\NIC}{NIC, DESY, Platanenallee 6, D-15738 Zeuthen, Germany}

\title{Time windows of the muon HVP from twisted-mass lattice QCD}

\author[a,b]{C.~Alexandrou}
\author[b]{S.~Bacchio}
\author[c]{P.~Dimopoulos}
\author[b]{J.~Finkenrath}
\author[d]{R.~Frezzotti}
\author*[e]{G.~Gagliardi}
\author[f]{M.~Garofalo}
\author[a,b]{K.~Hadjiyiannakou}
\author[g]{B.~Kostrzewa}
\author[h]{K.~Jansen}
\author[i]{V.~Lubicz}
\author[f]{M.~Petschlies}
\author[e]{F.~Sanfilippo}
\author[e]{S.~Simula}
\author[f]{C.~Urbach}
\author[j]{U.~Wenger}

\affiliation[a]{\CyprusU}
\affiliation[b]{\CyprusI}
\affiliation[c]{\Parma}
\affiliation[d]{\Romadue} 
\affiliation[e]{\RomatreINFN}
\affiliation[f]{\hiskp}
\affiliation[g]{\hpca}
\affiliation[h]{\NIC}
\affiliation[i]{\Romatre}
\affiliation[j]{\Bern}

\emailAdd{giuseppe.gagliardi@roma3.infn.it}

\abstract{
We present a lattice determination of the leading-order hadronic vacuum polarization (HVP) contribution to the muon anomalous magnetic moment, $a_{\mu}^{\rm HVP}$, in the so-called short and intermediate time-distance windows, $a_{\mu}^{\rm SD}$ and $a_{\mu}^{\rm W}$. We employ gauge ensembles produced by the Extended Twisted Mass Collaboration (ETMC) with $N_f = 2 + 1 + 1$ flavours of Wilson-clover twisted-mass quarks with masses of all the dynamical quark flavours tuned close to their physical values. The simulations are carried out at three values of the lattice spacing equal to $\simeq 0.057, 0.068$ and $0.080$ fm with spatial lattice sizes up to $L \simeq 7.6$~fm. For the short distance window we obtain $a_\mu^{\rm SD} = 69.27\,(34) \cdot 10^{-10}$, in agreement with the dispersive determination based on experimental $e^+ e^-$ data. For the intermediate window we get instead $a_\mu^{\rm W} = 236.3\,(1.3) \cdot 10^{-10}$, which is consistent with recent determinations by other lattice collaborations, but disagrees with the dispersive determination at the level of $3.6\,\sigma$.

}

\FullConference{%
  The 39th International Symposium on Lattice Field Theory (Lattice2022),\\
  8-13 August, 2022 \\
  Bonn, Germany 
}

\begin{document}
\maketitle

\section{Introduction}
The anomalous magnetic moment of the muon $a_{\mu} \equiv (g-2)/2$, is one of the most  precisely determined quantities in  physics, both experimentally and theoretically. It is a crucial quantity  for which a long-standing tension between the experimental value and the Standard Model (SM) prediction  might provide  important evidence for New Physics (NP) beyond the SM. From the theoretical side, the dominant source of uncertainty in the determination of $a_{\mu}$ comes from the leading-order (LO) Hadronic Vacuum Polarization (HVP) term $a_{\mu}^{\rm HVP}$ of order $\mathcal{O}(\alpha_{em}^{2})$. The most precise prediction for the HVP contribution has been obtained till now using a data-driven approach, in which the HVP contribution is reconstructed from the experimental cross section data for electron-positron annihilation into hadrons (the R-ratio method), using dispersion relations\,\cite{Davier:2017zfy, Keshavarzi:2018mgv, Colangelo:2018mtw, Hoferichter:2019mqg, Keshavarzi:2019abf, Davier:2019can}. The SM prediction for $a_{\mu}$, obtained using the dispersive value for the LO-HVP~\cite{Aoyama:2020ynm}, shows a remarkable tension of $4.2\,\sigma$ with the experimental result\,\cite{Muong-2:2021ojo}. Having accurate lattice determinations of the LO-HVP, using the pure SM theory, becomes thus of crucial important in order to crosscheck the dispersive result, especially in view of the fact that the dispersive determination, being data-driven, could be affected by NP contaminations present in $e^+ e^-$ data (see Ref.~\cite{Alexandrou:2022tyn} for a first study of this point). \\

On the lattice, the LO-HVP contribution is typically evaluated employing the so-called time momentum representation\,\cite{Bernecker:2011gh} in which the LO-HVP is obtained as
\begin{equation}
    \label{eq:amu_HVP}
    a_{\mu}^{\rm{HVP}} = 2 \alpha_{em}^2 \int_0^\infty ~ dt \, t^2 \, K(m_\mu t) \,V(t) ~ , ~  
\end{equation}
where $t$ is the Euclidean time, $m_{\mu}$ the muon mass, and the kernel function $K(z)$ is defined as\footnote{The leptonic kernel $K(z)$ is proportional to $z^2$ at small values of $z$ and it goes to $1$ for $z \to \infty$.}
\begin{equation}
    \label{eq:kernel}
    K(z) = 2 \int_0^1 dy ( 1- y) \left[ 1 - j_0^2 \left(\frac{z}{2}\frac{y}{\sqrt{1 - y}} \right) \right]~,\qquad j_{0}(y) = \frac{\sin{(y)}}{y} ~ . ~
\end{equation}
The Euclidean vector correlator $V(t)$ at vanishing three-momentum is defined as
\begin{equation}
     \label{eq:VV}
     V(t) \equiv - \frac{1}{3} \sum_{i=1,2,3} \int d^3{x} ~ \langle J_i(\vec{x}_f, t_f) J_i(\vec{x}_i, t_i) \rangle 
\end{equation}
with $J_\mu(x)$ being the electromagnetic (em) current operator
\begin{equation}
      \label{eq:Jmu}
     J_\mu(x) \equiv \sum_{f = u, d, s, c, ...} q_f ~ \overline{\psi}_f(x) \gamma_\mu \psi_f(x) ~ 
\end{equation}
 and $q_f$ the electric charge for the quark flavour $f$ (in units of the absolute value of the electron charge). A breakthrough concerning the precision achieved in the evaluation of the HVP came from the recent lattice result by the BMW Collaboration\,\cite{Borsanyi:2020mff}. Their value of $a_{\mu}^{\rm HVP}$ turns out to differ from the dispersive one at the level of $2.1\sigma$, giving rise to a value for the total muon anomaly that is much closer to the experimental result. The BMW results triggered in the last few years a joint effort by the lattice QCD community, with many collaborations trying to provide independent determinations of $a_{\mu}^{\rm HVP}$ with sub-percent accuracy. In this respect, important benchmark quantities,  tailored to ease the comparison across different lattice determinations, and which enable to probe $e^+ e^-$ annihilation data in different center-of-mass energy regions, are the so-called   time windows introduced by the RBC/UKQCD Collaboration\,\cite{RBC:2018dos}. They are given by
 \begin{equation}
 \label{eq:win_def}
 a_{\mu}^{w} = 2 \alpha_{em}^2 \int_0^\infty ~ dt \, t^2 \, K(m_\mu t)\, \Theta^{w}(t) \,V(t)\, , \qquad \{ w = \rm{SD},\rm{W},\rm{LD}\}~, 
 \end{equation}
 where the time-modulating functions $\Theta^{w}$ are defined as
 \begin{equation}
      \Theta^{\rm SD}(t) \equiv 1 -  \frac{1}{1 + e^{- 2 (t - t_0) / \Delta}} ~ , ~ 
      \Theta^{\rm W}(t)  \equiv  \frac{1}{1 + e^{- 2 (t - t_0) / \Delta}} -  \frac{1}{1 + e^{- 2 (t - t_1) / \Delta}} ~ , ~    
      \Theta^{\rm LD}(t)  \equiv  \frac{1}{1 + e^{- 2 (t - t_1) / \Delta}} ~ ,  
 \end{equation}
 with $t_0 = 0.4 ~ \rm{fm} ~ , ~  t_1 = 1 ~ {\rm fm} ~ , ~  \Delta = 0.15~{\rm fm}$, and $a_{\mu}^{\rm{HVP}} = a_{\mu}^{\rm{SD}} + a_{\mu}^{\rm{W}} + a_{\mu}^{\rm{LD}}$. Each time window can be separately compared with the dispersive determination, since the modulating functions $\Theta^{w}(t)$ have analogous \textit{energy counterparts} which allow to probe the low ($w={\rm LD}$), intermediate ($w={\rm W})$ and high ($w={\rm SD}$) energy part of the $e^+ e^- \rightarrow$ hadrons differential cross section (see Sec. II of\,\cite{Alexandrou:2022amy} for a detailed discussion). In particular the short-distance ($a_{\mu}^{\rm SD}$) and intermediate ($a_{\mu}^{\rm W}$) windows offer the possibility of a high-precision comparison between the two approaches, since the lattice vector correlator is typically very precise for Euclidean times $t \lesssim 1~{\rm fm}$. In Ref.~\cite{Alexandrou:2022amy} we presented the results of our calculation of the short and intermediate time window contributions.  The scope of this proceedings is to provide a short description of the analysis and a summary of our main findings.  

\section{Lattice setup and computational strategy}
Our results are based on simulations of $N_{f}=2+1+1$ Wilson-clover twisted-mass fermions\,~\cite{Frezzotti:2000nk} with all sea quark masses tuned very close to their physical value. We decompose the isosymmetric QCD contribution to $a_{\mu}^{w}$ as 
\begin{equation}
a_{\mu}^{w} = a_{\mu}^{w}(\ell) + a_{\mu}^{w}(s) + a_{\mu}^{w}(c) + a_{\mu}^{w}(\rm{disc.})~,
\end{equation}
where the first three terms correspond to the quark-connected light ($\ell=u+d$), strange ($s$) and charm ($c$) contributions, while $a_{\mu}^{w}({\rm disc.})$ corresponds to the sum of all quark-disconnected diagrams (flavour diagonal+off diagonal). Our determination of $a_{\mu}^{w}(\rm{disc.})$ is thoroughly discussed in Ref.~\cite{Alexandrou:2022osy}, here we focus on the evaluation of the    quark-connected contributions.  For this calculation, we produced four ensembles almost at the physical point and at three values of the lattice spacing in the range $0.056-0.08~{\rm fm}$.\footnote{For the charm contributions, we also employ a fourth lattice spacing $a\sim 0.09~{\rm fm}$, using three ensembles with higher-than-physical pion masses $M_{\pi} \in [260, 365]~{\rm MeV}$. The presence of heavier-than-physical pions does not require a physical point extrapolation as the charm contribution is largely unsensitive to the value of the sea light-quark mass.} Essential information on these ensembles are collected in Table~\ref{tab:simudetails}. Our strategy to compute the connected contributions to $a_{\mu}^{\rm{W}}$ and $a_{\mu}^{\rm{SD}}$ is based on the following steps:
\begin{itemize}
\item Usage of two different discretized versions of the local em current, peculiar to our twisted-mass LQCD setup, that in the following will be indicated as twisted-mass (``tm") and Osterwalder-Seiler (``OS")\,\cite{Frezzotti:2004wz}. The results obtained using the two currents only differ by $\mathcal{O}(a^{2})$ cut-off effects, enabling to approach the continuum limit in two different ways. 
\item The continuum limit extrapolation is performed at a fixed spatial volume $V= L_{\rm{ref}}^{3}$, with $L_{\rm{ref}}= 5.46~{\rm fm}$, corresponding to the volume of our two finest lattice spacing ensembles. At $\beta=1.778$, a smooth interpolation in $e^{-M_\pi L}$ of the results obtained on the cB211.072.64 and cB211.072.96 ensembles is performed. 
\item The valence strange- and charm-quark mass is tuned alternatively using two different hadronic inputs: the mass of the pseudoscalar $\eta_{ss'}$ and $\eta_c$, and that of the $\phi$ and $J/\Psi$ vector meson. 
\item The small mistuning of the pion mass w.r.t. its isoQCD value $M_{\pi}^{\rm{isoQCD}}= 135.0(2)~{\rm MeV}$ (see Table\,\ref{tab:simudetails}) is cured, for each gauge ensemble, through a first-principle evaluation of the corrections in both valence and sea sector. In the valence, the correction is evaluated explicitly by performing additional simulations at a slightly smaller value of the light-quark valence bare mass $a \mu_\ell$. Sea-quark mass mistuning effects are instead evaluated adopting the RM123 expansion method\,\cite{deDivitiis:2011eh, deDivitiis:2013xla, Giusti:2019xct} (see Appendix A of Ref.\,\cite{Alexandrou:2022amy} for more details). Such corrections are dominated by the valence one and globally turn out to be negligible for all contributions but $a_{\mu}^{\rm W}(\ell)$, for which they correspond to a $1-2\,\sigma$ upwards shift of the results.  
\end{itemize}
To the continuum extrapolated data at $L=L_{\rm{ref}}$ we must apply a finite-size correction in order to obtain the infinite-volume results. For all contributions but $a_{\mu}^{\rm W}(\ell)$, such corrections are expected and checked to be well within the errors, hence no correction is applied. For $a_{\mu}^{\rm W}(\ell)$  we apply instead a finite-size  correction $\Delta a_\mu^{\rm W}(\ell; L_{\rm{ref}})$ evaluated in the Meyer-Lellouch-L\"uscher-Gounaris-Sakurai (MLLGS) model\,\cite{Luscher:1985dn, Luscher:1986pf, Luscher:1990ux, Luscher:1991cf, Lellouch:2000pv, Meyer:2011um, Francis:2013fzp, Gounaris:1968mw} which assumes the dominance of the finite-size effects related to intermediate two-pion states and contains no free parameters. In what follows, our lattice data of $a_\mu^w(f)$ for $w = \{\rm{SD}, \rm{W} \}$ and $f = \{\ell, s, c \}$ are already interpolated at the physical pion mass $M_\pi^{phys} = M_\pi^{\rm{isoQCD}} = 135.0\,(2)$ MeV and at $L_{\rm{ref}} = 5.46$ fm.

\begin{table}[htb!]
\begin{center}
    \begin{tabular}{||c||c|c|c|c|c||c||}
    \hline
    ~~~ ensemble ~~~ & ~~~ $\beta$ ~~~ & ~~~ $V/a^{4}$ ~~~ & ~~~ $a$ (fm) ~~~ & ~~~ $a\mu_{\ell}$ ~~~ & ~ $M_{\pi}$ (MeV) ~ & ~ $L$ (fm)  ~ \\
  \hline
  cB211.072.64 & $1.778$ & $64^{3}\cdot 128$ & $0.07957~(13)$ & $0.00072$ & $140.2~(0.2)$ & $5.09$  \\
  
  cB211.072.96 & $1.778$ & $96^{3}\cdot 192$ & $0.07957~(13)$ & $0.00072$ & $140.1~(0.2)$ & $7.64$ \\
  
  cC211.060.80 & $1.836$ & $80^{3}\cdot 160$ & $0.06821~(13)$ & $0.00060$ & $136.7~(0.2)$ & $5.46$ \\
  
  cD211.054.96 & $1.900$ & $96^{3}\cdot 192$ & $0.05692~(12)$ & $0.00054$ & $140.8~(0.2)$ & $5.46$ \\
  \hline
    \end{tabular}
\end{center}
\caption{\it \small Parameters of the ETMC ensembles used in this work. We give the light-quark bare mass, $a \mu_\ell = a \mu_u = a \mu_d$, the lattice spacing $a$,  the pion mass $M_\pi$, and the lattice size $L$. }
\label{tab:simudetails}
\end{table}

\section{Short-distance window contributions}
In the short-distance window, only small Euclidean times of order $\mathcal{O}(t_{0})$ are relevant, since the contributions from times $t > t_{0}$ are exponentially suppressed by $\Theta^{\rm SD}(t)$. In this region of time, our correlator $V(t)$ is particularly precise, with the relative uncertainties being of order $\mathcal{O}(0.1\%)$ or smaller, for all flavour contributions. One of the main challenges in the determination of $a_{\mu}^{\rm SD}$ is represented by the continuum extrapolation, due to the presence of log-enhanced cut-off effects of order $\mathcal{O}(a^{2}\log{a})$ which are generated by the integration in the region of times $t$ of order $\mathcal{O}(a)$ (see Refs.\,\cite{DellaMorte:2008xb,Ce:2021xgd, Alexandrou:2022amy} for details). Such discretization effects, containing a positive power of the logarithm, are dangerous, since they slow down the convergence with respect to a pure $a^2$-scaling and may not be visible unless simulations at very small lattice spacing are performed. Since such artifacts are already generated in the free theory, we remove them explicitly by subtracting from our raw data for $a_{\mu}^{\rm SD}(f=\ell,s,c)$ the free-theory lattice artifacts evaluated using the same bare quark masses adopted in our numerical simulations for the different flavours. \\

\begin{figure}
\begin{center}
\includegraphics[scale=0.27]{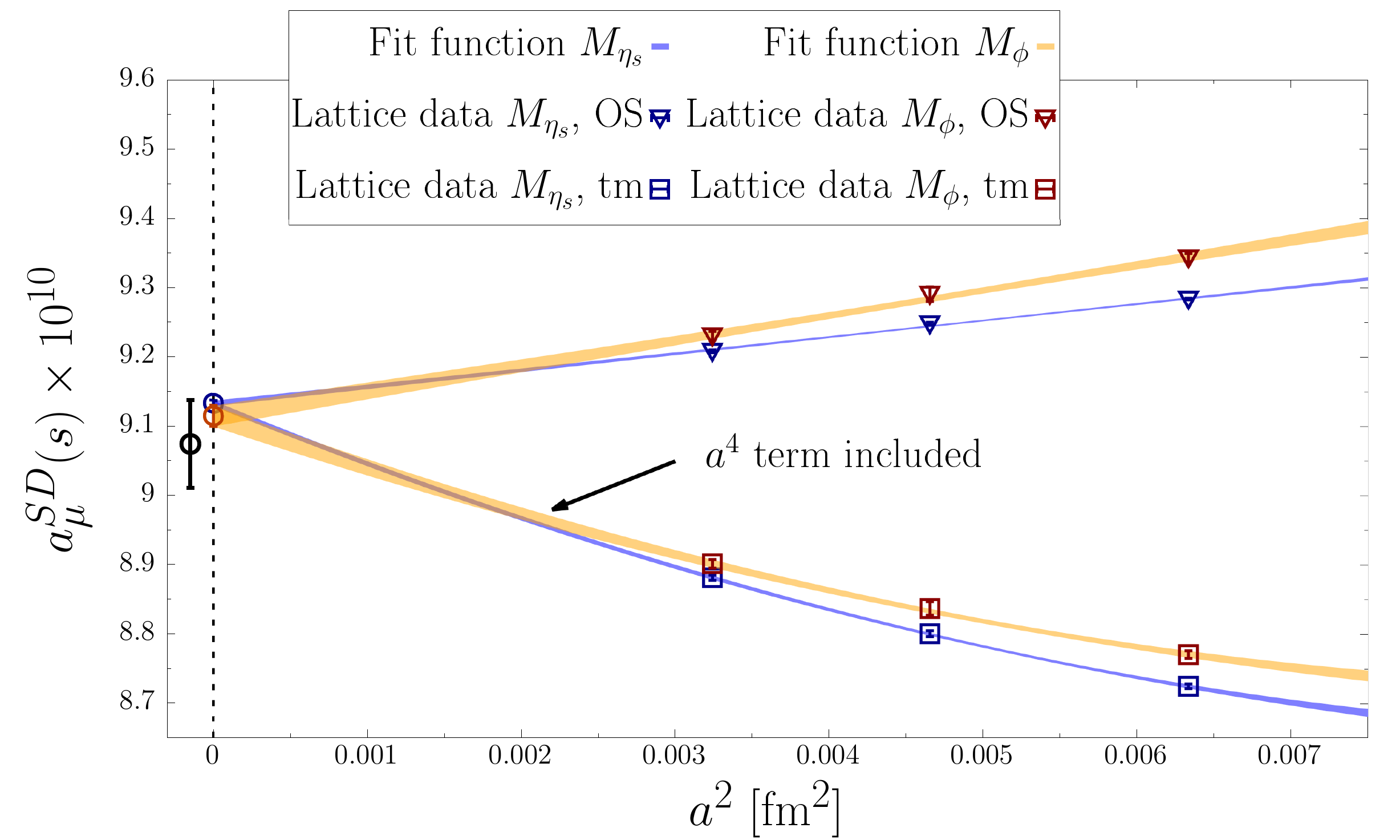}
\includegraphics[scale=0.27]{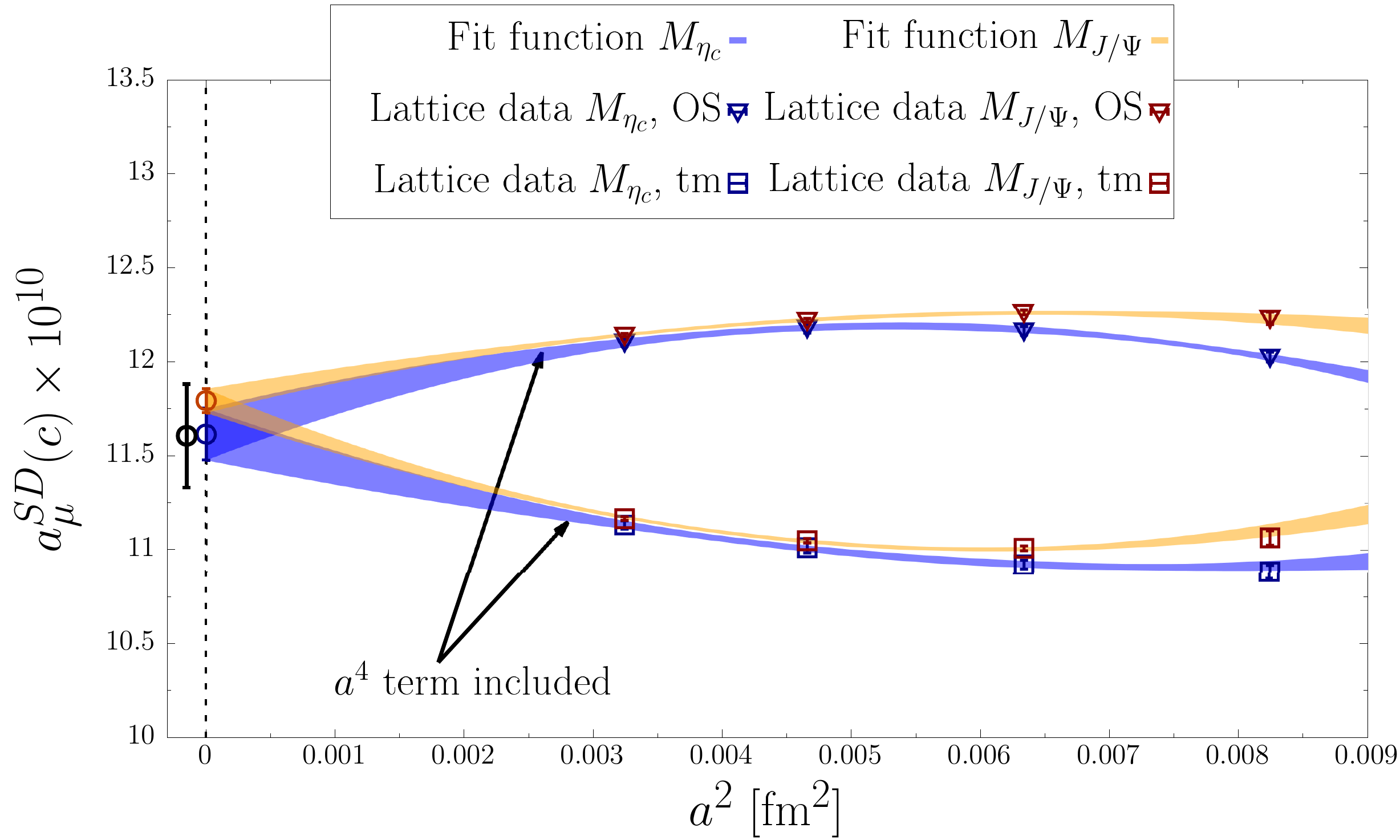}\\
\includegraphics[scale=0.25]{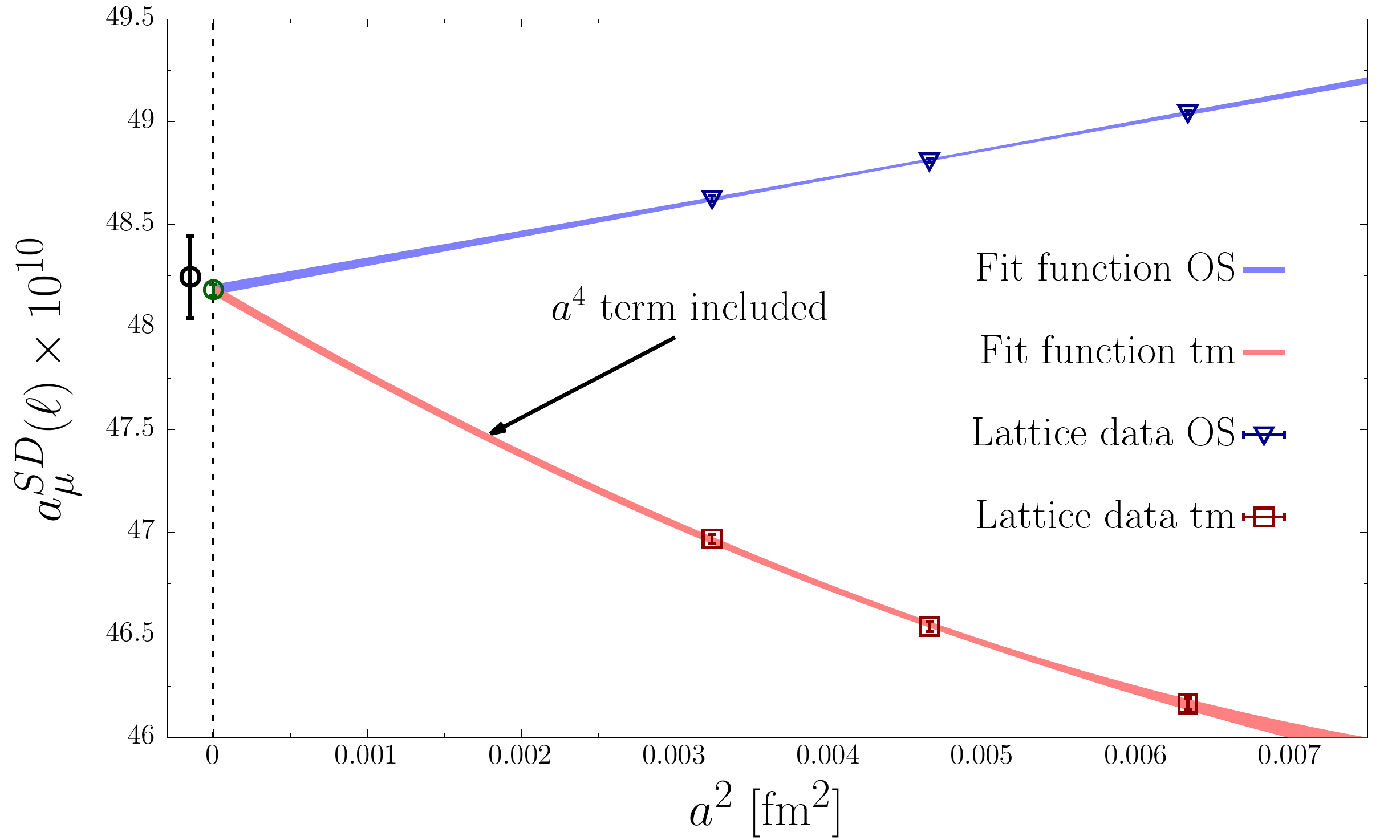}
\vspace{-0.3cm}
\caption{\it \small The light-quark (bottom), strange-quark (top-left) and charm-quark (top-right) connected contributions to the short-distance window $a_\mu^{\rm SD}$ versus the squared lattice spacing $a^2$ in physical units using both the ``tm" (triangles) and ``OS" (squares) regularizations. For the strange and charm contributions, the blue and red points correspond to the lattice data obtained using the masses of the $\eta_s$ ($\eta_c$) and $\phi$ ($J/\Psi$) mesons to obtain the physical strange (charm) quark mass. The solid lines correspond to the results of a representative (polynomial in $a^{2}$) combined continuum extrapolation. The black data points at $a^2 < 0$ correspond to our final results given by Eqs.\,(\ref{eq:amuSD_ell_final})-(\ref{eq:amuSD_charm_final}).}
\label{fig:SD_cont_lim}
\end{center}
\end{figure}

\begin{figure}[htb!]
\begin{center}
\includegraphics[scale=0.25]{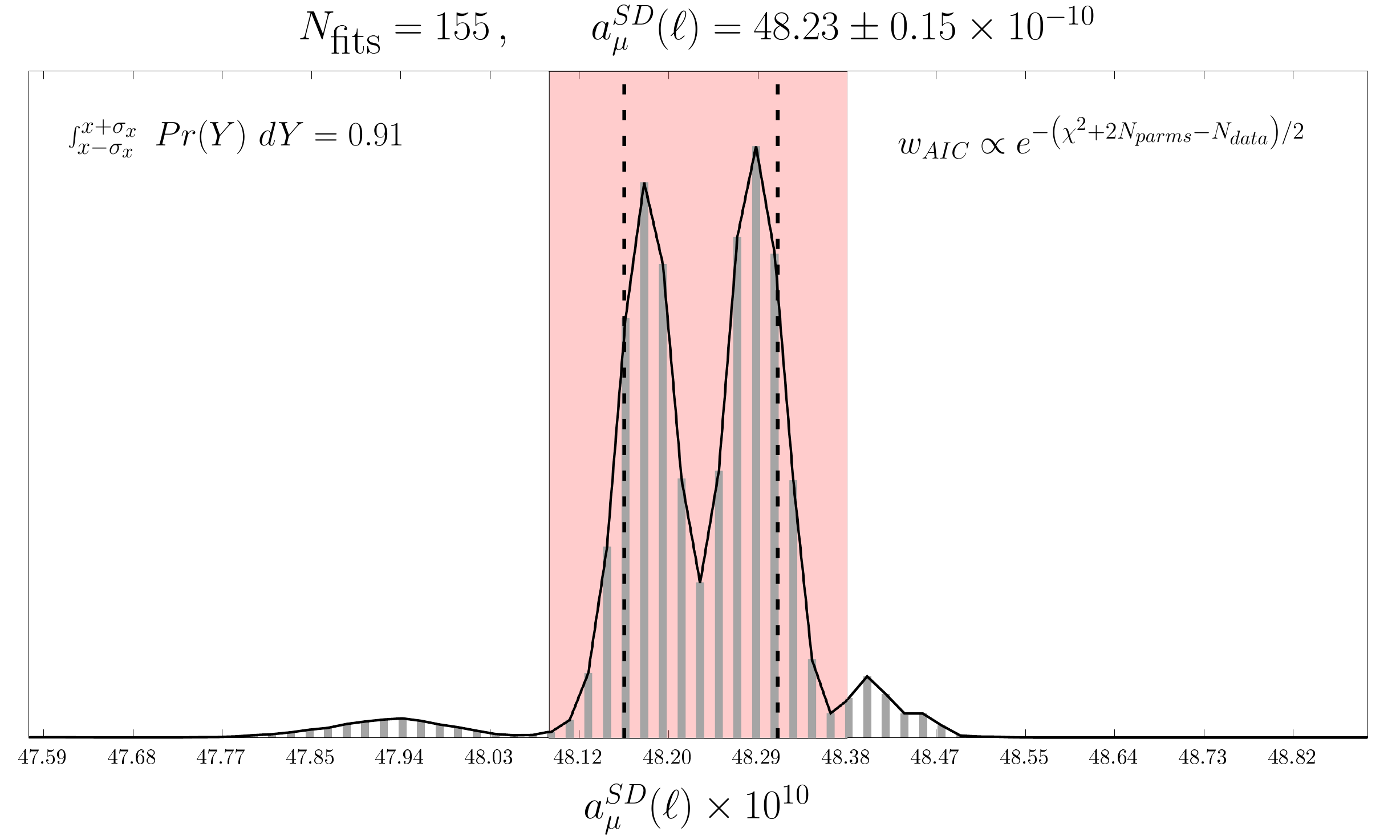} ~ 
\includegraphics[scale=0.25]{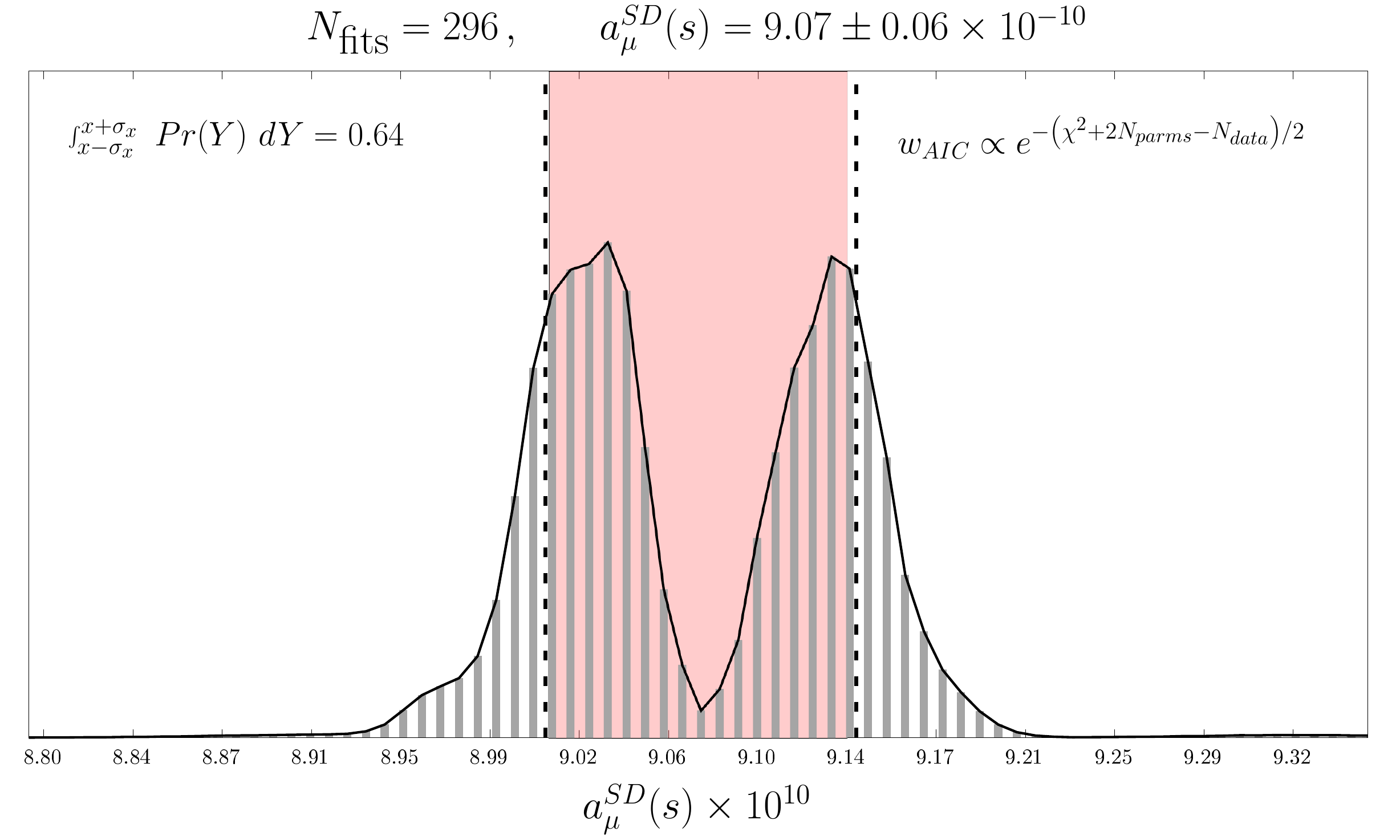} \\
\includegraphics[scale=0.25]{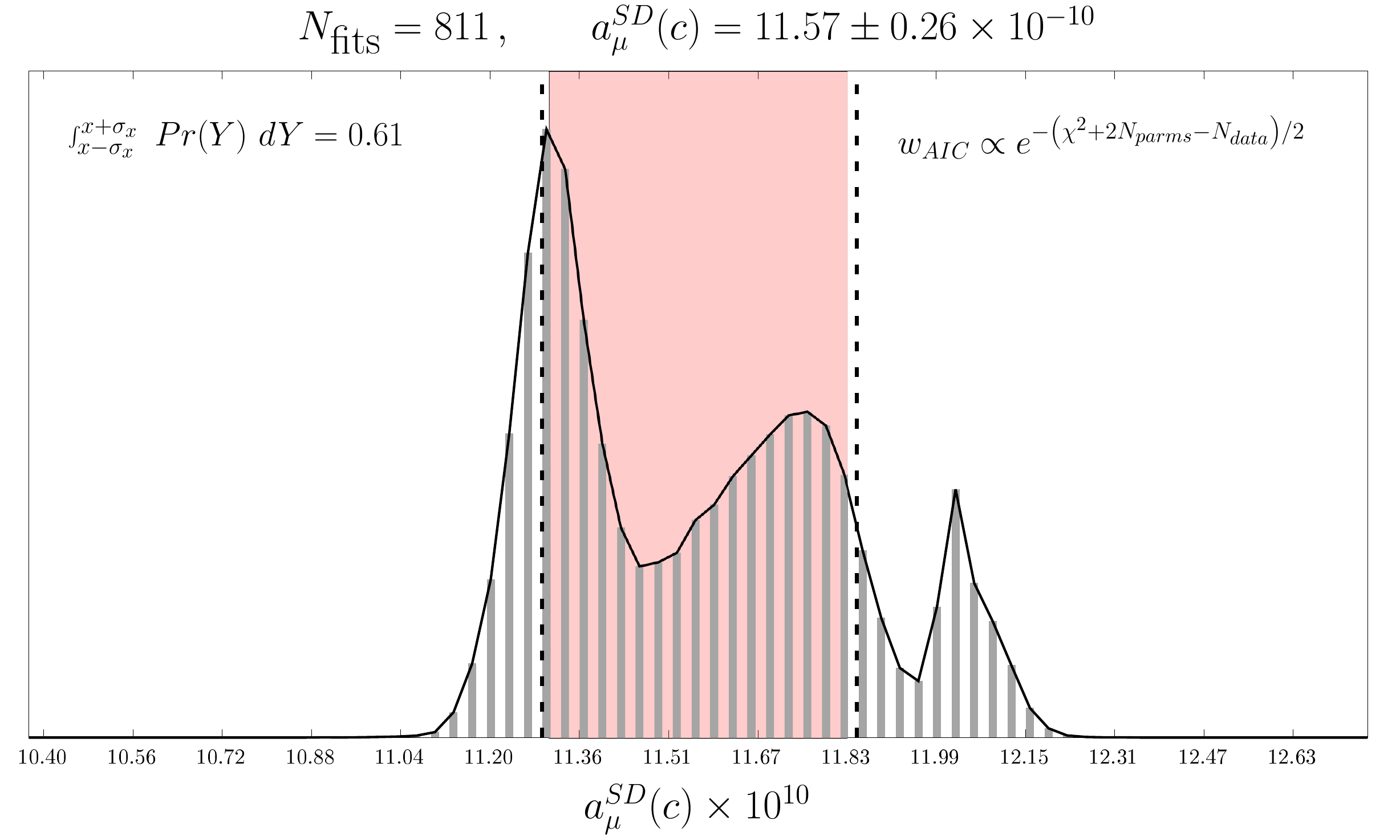}  
\vspace{-0.3cm}
\caption{\it \small Histograms of the results obtained adopting the AIC weights. For each contribution we show the number of fits, the average ($x$) and the error ($\sigma_x$) evaluated according to Eq.\,(\ref{eq:averaging}) and the cumulative probability corresponding to the interval $[x - \sigma_x, x + \sigma_x]$ given by the pink band. The vertical short-dashed lines correspond to the 16-th and 84-th percentiles of the p.d.f.~$Pr(Y)$.}
\label{fig:amuSD}
\end{center}
\end{figure}

The values of $a_\mu^{\rm SD}(\ell)$, $a_\mu^{\rm SD}(s)$ and $a_\mu^{\rm SD}(c)$ obtained after the subtraction of the perturbative lattice artifacts are shown in Fig.\,\ref{fig:SD_cont_lim} for both the ``tm" and ``OS" regularizations, together with a representative continuum limit extrapolation. The extrapolations are always carried out by fitting simultaneously the data corresponding to the two regularizations ``tm" and ``OS", and constraining the continuum extrapolated value to be the same\footnote{ In all cases the $\chi^{2}$ function to be minimized has been constructed taking into account the correlation between the ``tm" and ``OS" data points corresponding to the same ensemble.}. A detailed description of the fit Ans{\"a}tze which have been considered can be found in Section III of Ref.\,\cite{Alexandrou:2022amy}. For each contribution we performed hundreds of continuum fits which have been  combined making use of the procedure developed in Ref.\,\cite{EuropeanTwistedMass:2014osg}: starting from $N$ fit results with mean values $x_k$ and uncertainties $\sigma_k$ ($k=1,\cdots,N$), the final average $x$ and uncertainty $\sigma_x$ are given by
\begin{equation}
    \label{eq:averaging}
    x = \sum_{k=1}^N \omega_k ~ x_k ~ , ~ \qquad
    \sigma_x^2 = \sum_{k=1}^N \omega_k ~ \sigma_k^2 + \sum_{k=1}^N \omega_k ~ (x_k - x)^2 ~ , ~
\end{equation}
where $\omega_k$ represents the weight associated with the $k$-th fit. 
We have excluded from the average all fits having $d.o.f. = 1$ in order to avoid overfitting. Then, we have considered two choices for the weights $\omega_k$. The first one is based on the Akaike Information Criterion (AIC)\,\cite{Akaike}, namely 
\begin{equation}
    \label{eq:AIC}
    \omega_k \propto e^{- (\chi_k^2 + 2 N_{parms} - N_{data}) / 2} ~ , ~
\end{equation}
where $\chi_k^2$ is the value of the $\chi^2$ variable for the $k$-th computation, $N_{parms}$ is the number of free parameters and $N_{data}$ the number of data points.
Since in our fits the number of d.o.f. is limited, we also tried a second choice for $\omega_k$ given by a step function 
\begin{equation}
   \label{eq:stepF}
   \omega_k \propto \Theta\left[ 1 + 2 \sqrt{ \frac{2}{d.o.f.}} - \frac{\chi_k^2}{d.o.f.} \right] ~ , ~ 
\end{equation}
where $1$ is the mean value and $\sqrt{2 / d.o.f.}$ is the standard deviation of the $\chi^2/d.o.f.$ distribution. 
The (typically small) difference between the results obtained with the above two choices of $\omega_k$, is added as a systematic error in the final error budget. We obtain
\begin{eqnarray}
    \label{eq:amuSD_ell_final}
    a_\mu^{\rm SD}(\ell) & = & 48.24 ~ (3)_{stat} ~ (20)_{syst} \cdot 10^{-10} = 48.24 ~ (20) \cdot 10^{-10} ~ , ~ \\[2mm]
    \label{eq:amuSD_strange_final}
    a_\mu^{\rm SD}(s) & = & 9.074 ~ (14)_{stat} ~ (62)_{syst} \cdot 10^{-10} = 9.074 ~ (64) \cdot 10^{-10} ~ , ~ \\[2mm]
    \label{eq:amuSD_charm_final}    
    a_\mu^{\rm SD}(c) & = & 11.61 ~ (9)_{stat} ~ (25)_{syst} \cdot 10^{-10} = 11.61 ~ (27) \cdot 10^{-10} ~ . ~ 
\end{eqnarray}
In Fig.\,\ref{fig:amuSD} we show the distribution of the fit results corresponding to the AIC weights $\omega_{k}$.


\section{Intermediate window contributions}
\begin{figure}
\begin{center}
\includegraphics[scale=0.27]{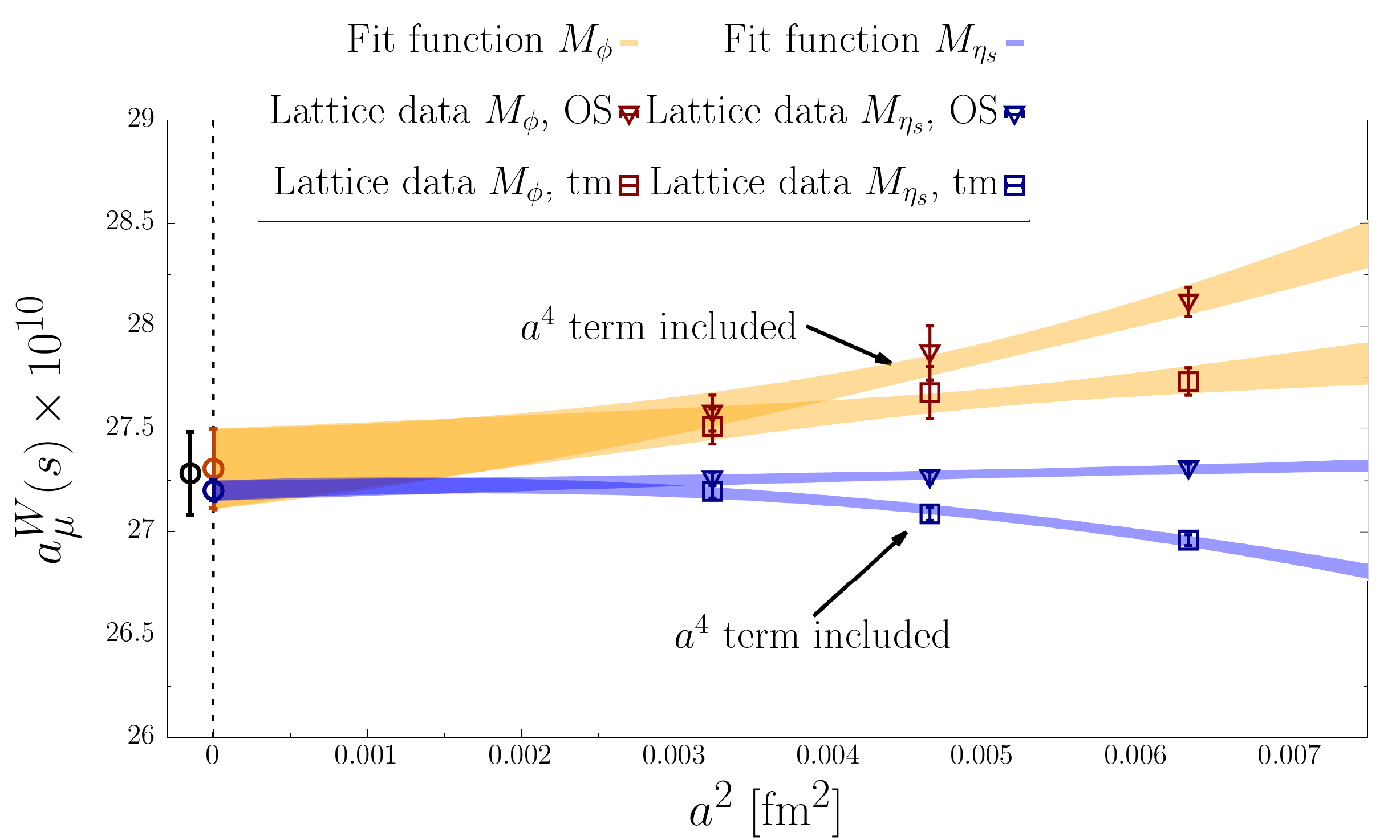}
\includegraphics[scale=0.27]{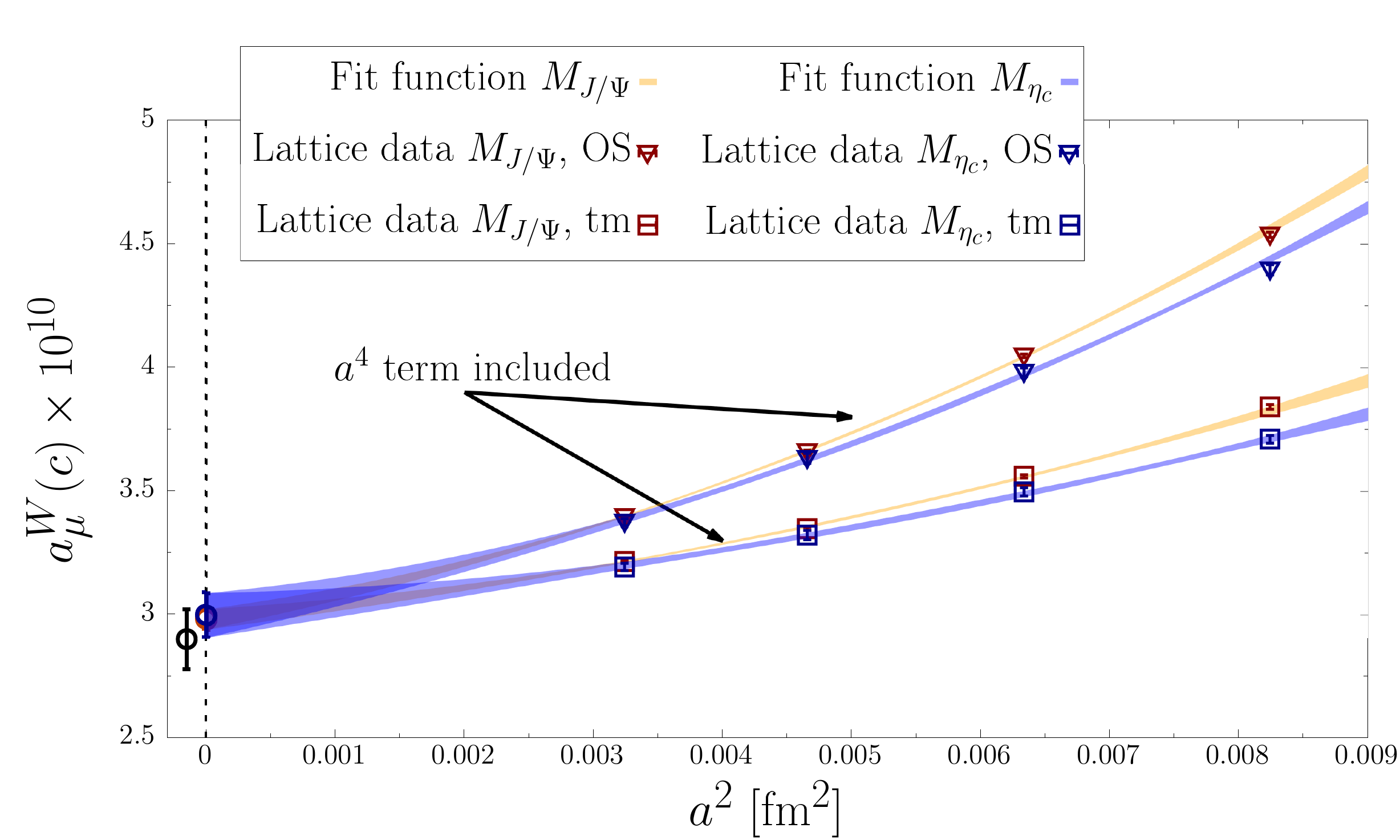}\\
\includegraphics[scale=0.25]{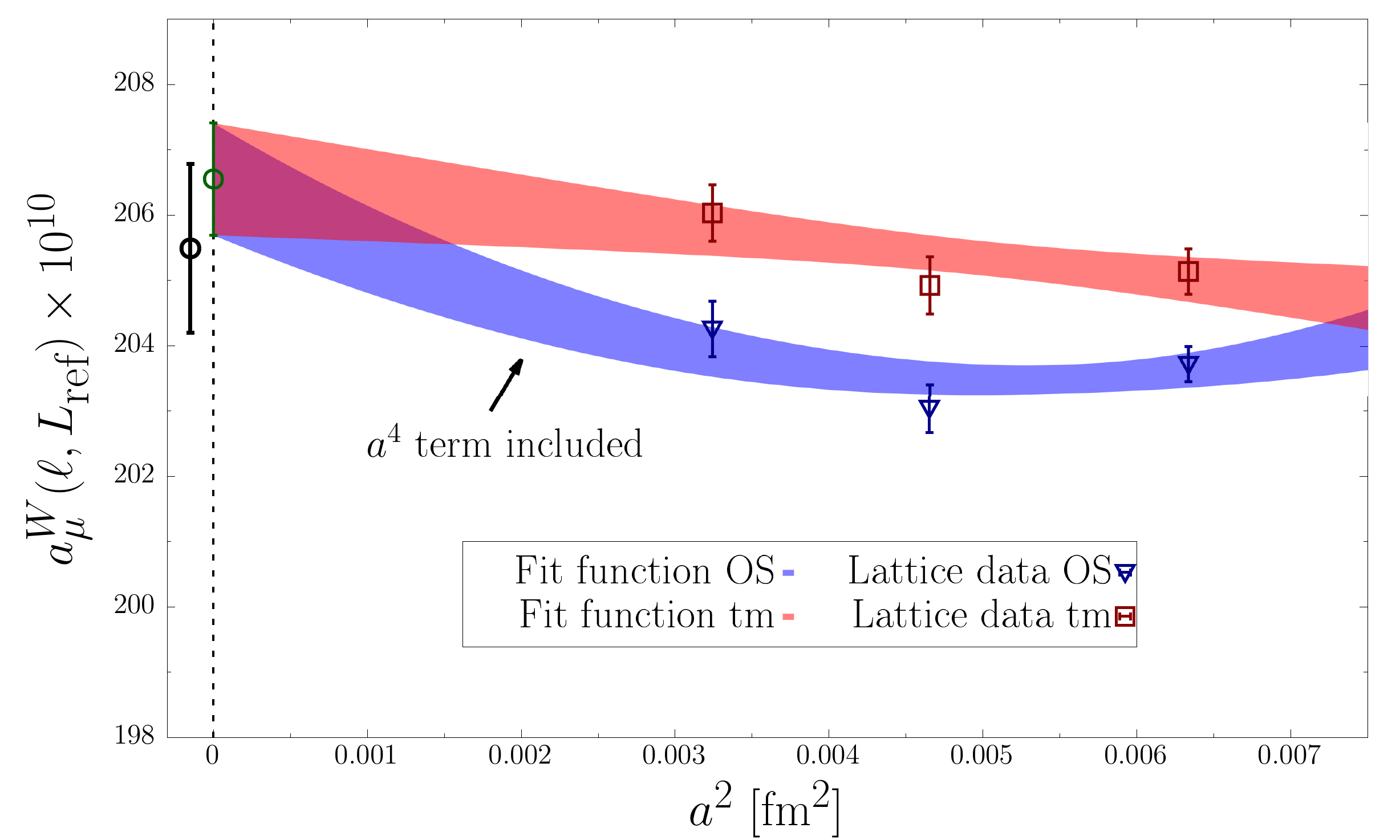}
\vspace{-0.3cm}
\caption{\it \small Same as in Fig.\,\ref{fig:SD_cont_lim} for the intermediate window contributions $a_{\mu}^{\rm W}(\ell), a_{\mu}^{\rm W}(s), a_{\mu}^{\rm W}(c)$.}
\label{fig:W_cont_lim}
\end{center}
\end{figure}

\begin{figure}[htb!]
\begin{center}
\includegraphics[scale=0.25]{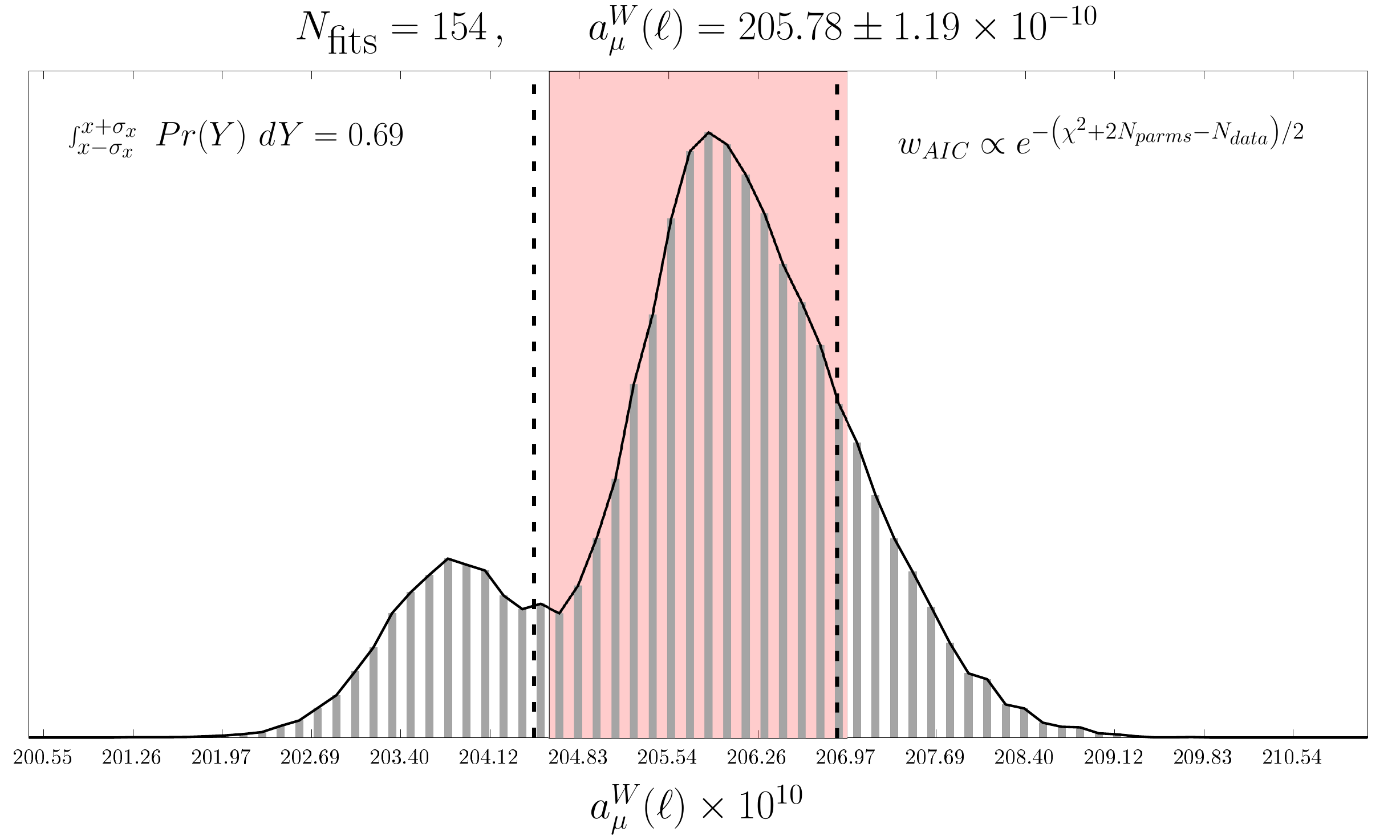} ~ 
\includegraphics[scale=0.25]{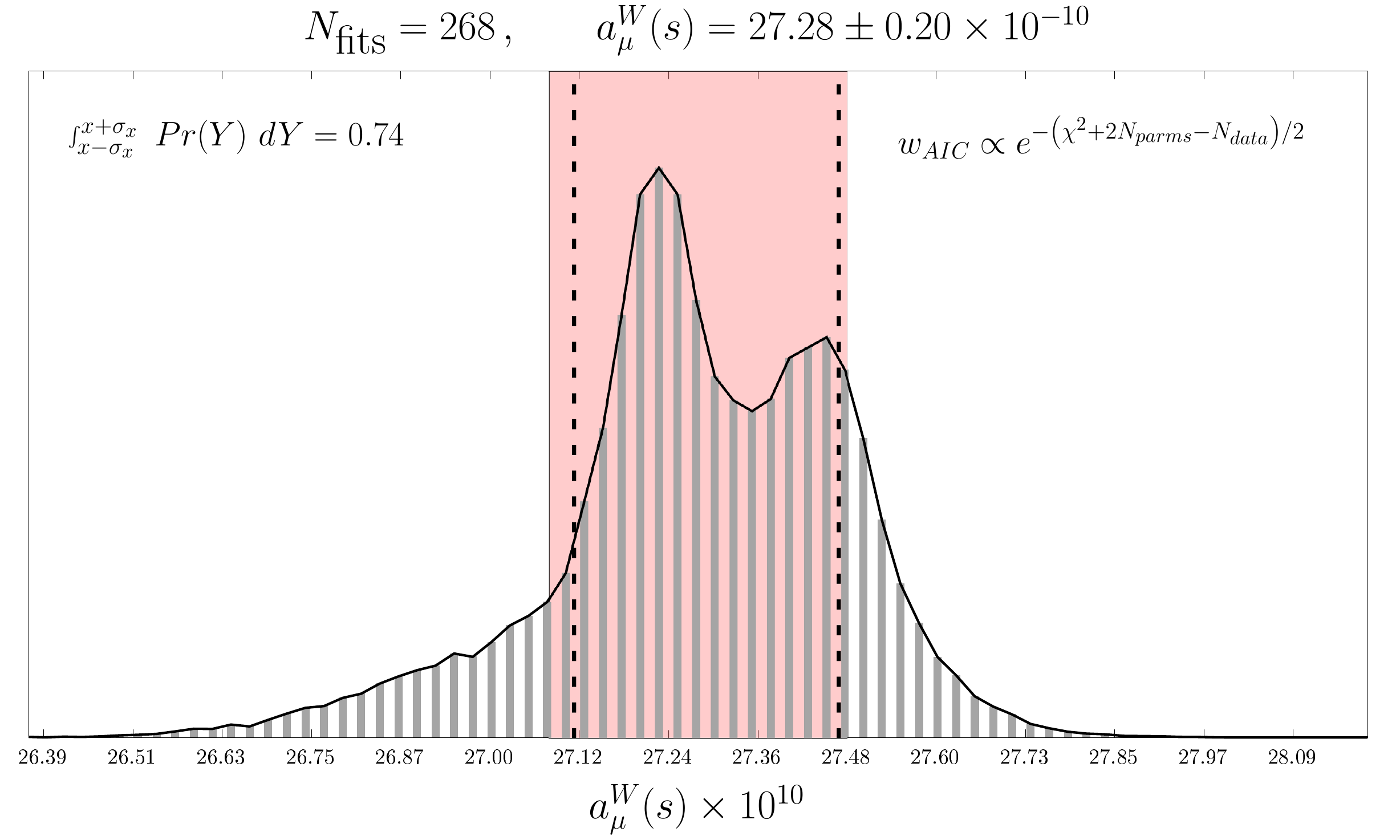} \\
\includegraphics[scale=0.25]{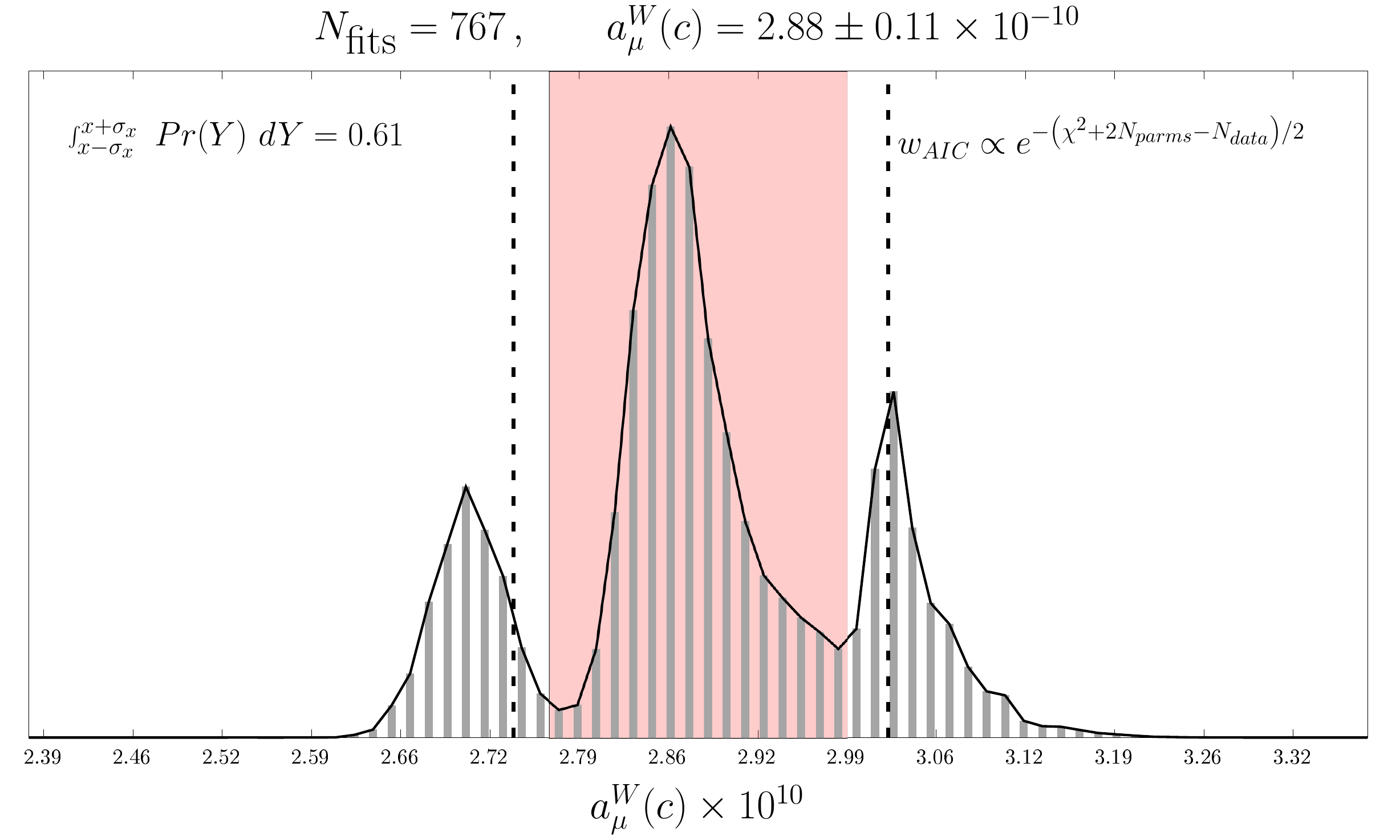}  
\vspace{-0.3cm}
\caption{\it \small Same as in Fig.\,\ref{fig:amuSD} for the intermediate window contributions $a_{\mu}^{\rm W}(\ell), a_{\mu}^{\rm W}(s), a_{\mu}^{\rm W}(c) $.   }
\label{fig:amuW}
\end{center}
\end{figure}

In the case of the intermediate window,  the accuracy of our lattice data is of order $\mathcal{O}(0.2\%)$ for $a_{\mu}^{\rm W}(\ell)$, of order $\mathcal{O}(0.2\%)$ ($\mathcal{O}(0.6\%)$) for $a_{\mu}^{\rm W}(s)$ when using the $\eta_{ss'}$ ($\phi$) mass to determine the physical strange-quark mass, while for $a_{\mu}^{\rm W}(c)$ the accuracy is typically of order $\mathcal{O}(0.5\%)$ ($\mathcal{O}(0.2\%)$) when using the $\eta_{c}$ ($J/\Psi$) mass to determine the physical charm-quark mass. In contrast to the short-distance window, there are no discretization effects of the type $a^2 \, \mbox{log}(a)$, thanks to the exponential suppression of the modulating function $\Theta^{\rm W}(t)$ at small values of $t \approx a$. Therefore, no subtraction of the perturbative lattice artifacts is carried out. Our lattice data corresponding to the ``tm" and ``OS" regularizations are shown in Fig.\,\ref{fig:W_cont_lim} together with a representative example of the combined continuum-limit extrapolation. As in the case of the short-distance window, we carry out hundreds of fits which are then combined using Eqs.\,(\ref{eq:averaging})-(\ref{eq:stepF}). In Fig.\,\ref{fig:amuW} we show the distribution of the fit results corresponding to the AIC weights $\omega_{k}$. Our final values are 
\begin{eqnarray}
    \label{eq:amuW_ell_Lref}
    a_\mu^{\rm W}(\ell, L_{\rm{ref}}) & = & 205.5 ~ (0.7)_{stat} ~ (1.1)_{syst} \cdot 10^{-10} = 205.5 ~ (1.3) \cdot 10^{-10} ~ , ~ \\[2mm]
    \label{eq:amuW_strange_final}
    a_\mu^{\rm W}(s) & = & 27.28 ~ (13)_{stat} ~ (15)_{syst} \cdot 10^{-10} = 27.28 ~ (20) \cdot 10^{-10} ~ , ~ \\[2mm]
    \label{eq:amuW_charm_final}    
    a_\mu^{\rm W}(c) & = & 2.90 ~ (3)_{stat} ~ (12)_{syst} \cdot 10^{-10} = 2.90 ~ (12) \cdot 10^{-10} ~ . ~ 
\end{eqnarray}
To the result of Eq.\,(\ref{eq:amuW_ell_Lref}) we add the volume correction $-\Delta a_\mu^{\rm W}(\ell, L_{\rm{ref}}) = 1.00 ~ (20) \cdot 10^{-10}$ evaluated within the MLLGS model (see Appendix F of Ref.\,\cite{Alexandrou:2022amy}) which leads to $a_{\mu}^{\rm W}(\ell) =  206.5 ~ (1.3) \cdot 10^{-10}$.

\section{ Comparison with data-driven determinations and conclusions}
In order to compare with the dispersive results, we must add to our $N_{f}=2+1+1$ isoQCD contributions
\begin{eqnarray}
a_{\mu}^{\rm SD}(\ell,s,c,{\rm disc.}) & = & \left[ 48.24~(20), 9.074~(64), 11.61~(27) , - 0.06~(5)         \right] \cdot 10^{-10} ~, \\
a_{\mu}^{\rm W}(\ell,s,c,{\rm disc.}) & = & \left[ 206.5(1.3), 27.28~(20), 2.90~(12), - 0.78~(21)         \right] \cdot 10^{-10} ~,
\end{eqnarray}
the (tiny) QED and strong isospin-breaking corrections to $a_{\mu}^{\rm W}$, which we take from BMW'20\,\cite{Borsanyi:2020mff} ($a_{\mu}^{\rm W}({\rm QED+SIB}) = 0.43(4) \cdot 10^{-10}$)
and the $b-$quark and QED contributions to $a_{\mu}^{\rm SD}$, which we evaluate  using the rhad software package\,\cite{Harlander:2002ur} ($a_{\mu}^{\rm SD}(b+\rm{QED}) = 0.35 \cdot 10^{-10}$). We obtain:
\begin{equation}
\,\,\,\,\, a_{\mu}^{\rm SD}(\rm{ETMC}) = 69.27~(34) \cdot 10^{-10}\, , \qquad a_{\mu}^{\rm W}(\rm{ETMC}) = 236.3~(1.3) \cdot 10^{-10}~,
\end{equation}
to be compared with the dispersive determinations\,\cite{Colangelo:2022vok}
\begin{equation}
\,\,\,\,\quad a_{\mu}^{\rm SD}(e^+ e^- ) = 68.4~(5)  \cdot 10^{-10}\, , \quad \,\,\, \qquad a_{\mu}^{\rm W}(e^+ e^- ) = 229.4 ~(1.4) \cdot 10^{-10}~.
\end{equation}
Our result for $a_{\mu}^{\rm SD}$ is in agreement at the level of $1.4\, \sigma $ with $e^+ e^-$ data, while for $a_{\mu}^{\rm W}$ we observe a discrepancy of $3.6\,\sigma$. Our findings for the different flavour contributions to $a_{\mu}^{\rm W}$ are also in remarkable good agreement with those
from other lattice groups (see\,\cite{Alexandrou:2022amy} for details). For the full $a_{\mu}^{\rm W}$, our result turns out to be in excellent agreement with its analog in the BMW'20\,\cite{Borsanyi:2020mff} ($a_\mu^{\rm W}({\rm BMW}) = 236.7(1.4) \cdot 10^{-10}$) and CLS'22\,\cite{Ce:2022kxy} ($a_\mu^{\rm W}({\rm CLS}) = 237.30\,(1.46) \cdot 10^{-10}$) papers. In conclusion, our (first) lattice determination of $a_{\mu}^{\rm SD}$ shows that the high-energy part of the $e^+ e^- \rightarrow$ hadron differential cross-section is in agreement with SM predictions. However, our determination of $a_{\mu}^{\rm W}$, which is in line with the results obtained by other lattice groups, points in the direction of a severe discrepancy w.r.t. the dispersive value (the discrepancy grows to $4.5\,\sigma$ if we average ETMC'22, BMW'20 and CLS'22 results), which definitely deserves further investigations.

\section*{Acknowledgments}
We thank all members of ETMC for the most enjoyable collaboration. We are very grateful
to G. Martinelli and G.C. Rossi for many discussions on the lattice setup and the methods
employed in this work. We thank N. Tantalo for valuable discussions about the physical
information that can be obtained by comparing experimental data on $e^{+}e^{-} \rightarrow$ hadrons with SM lattice predictions for observables related to the photon HVP term.

{
\bibliography{biblio}
\bibliographystyle{JHEP}
}

\end{document}